\documentstyle[11pt,newpasp,twoside, epsf]{article}
\def\edcomment#1{\iffalse\marginpar{\raggedright\sl#1\/}\else\relax\fi}
\reversemarginpar

\begin{document}
\title{The Spectroscopic Orbital Period of HD 45166}

\author{Alexandre S. Oliveira}

\affil{Instituto de Astronomia, Geof\'{\i}sica e Ci\^encias Atmosf\'ericas, USP, S\~ao Paulo, CP9638, Brazil}

\author{J. E. Steiner}

\affil{Minist\'erio da Ci\^encia e Tecnologia, DF, Brazil}

\begin{abstract}
HD 45166 was selected as a candidate to the V Sagittae Class.
It's spectrum shows quite narrow emission lines of He, C, N, and O, possibly indicating a low inclination
for the binary system. We performed high resolution and high stability Coud\'e spectroscopy
in order to search for low amplitude radial velocity variations. The orbital period determined from
the radial velocity curve is 0.357 days, and the radial velocity semi amplitude is
K=3.8 km/s. The system inclination, determined from the mass function of the secondary star, is
 $0.7\deg < i < 1.0\deg$.
\end{abstract}

\section{Introduction}

The class of the V Sagittae stars (Steiner \& Diaz 1998),
is composed of galactic binary systems with similar observational
properties. The stars of this class are spectroscopically characterized by the simultaneous presence of the
emission lines of OVI and NV and by the strength of the HeII 4686{\AA} emission line,
usually more than 2 times stronger than H$\beta$. The orbital periods are distributed between 5 and 12 hr.
The V Sagittae stars are possible galactic
counterparts of the close binary supersoft X-Ray sources (CBSS), found in
the Magellanic Clouds and in M31 (Steiner \& Diaz 1998).
The CBSS are presently accepted to be binary systems containing massive white dwarfs with stable nuclear burning of
accreted matter on its surface (van den Heuvel et al. 1992). This stable nuclear burning can occur when the mass
transfer is very high ($10^{-7} M_{\sun}$~ yr$^{-1}$), a situation found in systems with mass ratios
($q = M_2/M_1$) inverted when compared
 to the mass ratios usually found in cataclysmic variables (Kahabka \& van den Heuvel 1997).

HD 45166 is classified
in the Sixth Catalogue of Galactic Wolf-Rayet Stars (van der Hucht et al. 1981) as a low mass
WR-like star of spectral type qWR + B8V.
However, its spectrum shows unusually narrow emission
 lines for a Wolf-Rayet star, besides a solar chemical composition and both WN and WC features.
Willis \& Stickland (1983) confirmed the binary nature of the system
with the observation of a UV photospheric absorption spectrum and reported the evidence for variability in the
strengths of the NIV, NV, CIV and HeII emission lines, probably arising
in structural changes in a wind.
A SdOp rather than qWR interpretation is suggested,
where the "p" designation indicates the narrow emission line spectrum. These authors report that no velocity variability
in excess of $\sim 10$ km/s was found in the optical spectra of HD 45166.
If the amplitude of radial velocity variations
is lower than 10 km/s this could mean a low inclination for the binary system, which is compatible with the narrow emission
lines found in its spectra.

In an effort to determine the orbital period of HD 45166 we performed high resolution and high stability
Coud\'e spectroscopy at the 1.6 m telescope
of the Laborat\'orio Nacional de Astrof{\'{\i}}sica, Brazil. A total of 42 spectra
were obtained with the 1800 l/mm dispersion grating, resulting in
0.2{\AA} FWHM spectral resolution.
Typical rms residuals in wavelength calibrations are $\sim 2$ m{\AA}.

\section{The spectroscopic period}

Radial velocities were measured with the cross-correlation method using as template the average
spectrum. A period search routine based on Fourier analysis with cleaning of spectral windows effects (Roberts et al. 1987)
was applied to the RV data.
The resultant power spectrum clearly shows a signal at $P = 0.357$ d.
The ephemeris for this period is

\begin{center}

$T_{0(spec)}\mathrm{(HJD)} = 2450914.484 {\pm}0.030 + 0.357 {\pm} 0.009  E$

\end{center}

where the zero phase is defined as the crossing from positive to negative values when
compared to $\gamma$. The sinusoid fit to the radial velocity curve yields the values of
$\gamma = - 0.8$ km/s to the systemic velocity and K =
$3.8 $ km/s to the radial velocity semi amplitude.

Assuming that the secondary star is at the main sequence and fills its Roche Lobe we can use the mean empirical mass-period
relationship (Warner 1995) to derive a mass of $M_2=0.95 M_{\sun}$. The
primary star, being a white dwarf with hydrostatic nuclear burning,
requires a mass larger than $0.5 M_{\sun}$ (Hachisu et al. 1999) but smaller than the
secondary star ($0.50 < M_1 < 0.95$) in order to keep the inverted mass ratio. With the observed
period, radial velocity amplitude and the standard relationship for the mass function of the secondary star,
the inclination of the system would be quite low,
$0.7\deg < i < 1.0\deg$.


\begin{references}
\reference Hachisu, I., Kato, M., Nomoto, K., Umeda, H. 1999, ApJ, 519, 314
\reference Kahabka, P., van den Heuvel, E.P.J. 1997, ARA\&A, 35, 69
\reference Roberts, D.H., Lehar, J., Dreher, J.W. 1987, AJ, 93, 968
\reference Steiner, J.E., Diaz, M.P. 1998, PASP, 110, 276
\reference van den Heuvel, E.P.J., Bhattacharya, D., Nomoto, K., Rappaport, S.A. 1992, A\&A, 262, 97
\reference van der Hucht, K.A., Conti, P.S., Lundstrom, I., Stenholm, B. 1981, Space Sci. Rev., 28, 227
\reference Warner, B. 1995, in Cataclysmic Variable Stars, Cambridge University Press, Cambridge
\reference Willis, A.J., Stickland, D.J. 1983, MNRAS, 203, 619	
	
	
\end{references}
\end{document}